\documentclass[11pt,a4paper]{article}
\usepackage{amsfonts,amssymb,amsmath,amsthm,cite}
\usepackage{graphicx}
\usepackage{slashed}

\begin{document}
\title{Note on correlation functions in conformal quantum mechanics}
\author{Sadi Khodaee and Dmitri Vassilevich\\
CMCC - Universidade Federal do ABC, Santo Andr\'e, SP, Brazil}
\maketitle
\begin{abstract}
We suggest a method to compute the correlation functions in conformal quantum mechanics (CFT$_1$) for the fields that transform under a non-local representation of $\mathfrak{sl}(2)$ basing on the invariance properties. Explicit calculations of 2- and 3-point correlation functions are presented.
\end{abstract}
\section{Introduction}\label{sec:intro}
Conformal quantum mechanical models are being studied for more than four decades starting with the paper by de Alfaro, Fubini and Furlan (dAFF) \cite{deAlfaro:1976vlx}, see also \cite{Jackiw:1972cb}. There is very large literature on this subject, from which we like to mention the analysis of correlation functions in \cite{Chamon:2011xk,Jackiw:2012ur} and the supersymmetric extensions 
\cite{Fedoruk:2011aa,Khodaee:2012bg}.

The modern interest to conformal quantum mechanics is related to the AdS$_2$/CFT$_1$ correspondence. The general arguments in favor of conformal quantum mechanics as a dual of the AdS$_2$ gravity were presented in \cite{Chamon:2011xk}. The work \cite{Vassilevich:2013ai} pinpointed the quantum mechanical models that are boundary duals to generalized 2D gravities on a cylinder. A very interesting example of a quantum mechanical dual to a nearly AdS$_2$ gravity was studied in \cite{Maldacena:2016upp}. A useful test to check a conjectured holographic correspondence is to compare the correlation functions.

Fields in CFT$_1$ or in conformal quantum mechanics are functions of a real variable $t$. Primary fields transform under the conformal algebra $\mathfrak{sl}(2)$ according the standard local representation (when all generators are given by differential operators). Two- and three-point correlation functions of such fields are defined up to a constant factor by their invariance properties.

One may hope that the same property holds true also for the fields that transform under different (though equivalent) representations of $\mathfrak{sl}(2)$. However, technical implementation of this expectation may be quite involved. In this short note we study the fields which transform under a representation that has two generators unchanged, while the third one contains a non-local term. We show, that two-point functions may be computed by solving the invariance conditions. For three-point functions, these equations are too complicated. To find the correlation functions we use fractional integrals and fractional derivatives. Although the generators are only slightly modified, the form of the correlation functions changes considerably. 

\section{Non-local representations of $\mathfrak{sl}(2)$}\label{sec:rep}

The  algebra of $\mathfrak{sl}(2)$ has three generators $E^i$, $i\in\{-,0,+\}$, satisfying the following commutation relations:
\begin{eqnarray}\label{algeb-sl2}
\nonumber \left[E^+,  E^-\right] &=& 2 i \ E^0 ,\\ \nonumber
\left[E^+,  E^0\right]&=& i \ E^+,\\ 
\left[E^-,  E^0\right] &=& - i \ E^-.
\end{eqnarray}
The Casimir operator is given by $J^2=\frac{1}{2}(E^+ E^- +E^- E^+)- (E^0)^2$.
This algebra admits a representation in terms of the differential operators acting on functions on the real line
\begin{eqnarray}\label{standard-rep}
E^+ = H &=& -i \partial_t \\ \nonumber
E^0 = D &=& i (t\partial_t  + \lambda)\\ \nonumber
E^- = K &=& -i (t^2  \partial_t   +2\lambda  \ t).
\end{eqnarray}
For this representation,
\begin{equation}
J^2=C(\lambda)\equiv \lambda (\lambda -1).\label{Clambda}
\end{equation}

As usual, we define the conformal primaries as the fields which transform in the following way
\begin{equation}
\Phi(t) \ \rightarrow \Phi'(t)= \left( \frac{\partial t'}{\partial t} \right)^\lambda \ \Phi(t'),
\label{Phiprime}
\end{equation}
under the $\mathfrak{sl}(2)$ transformations
\begin{equation}
t\to t'=t+\omega_++\omega_0 t +\omega_+t^2\,,
\end{equation}
where $\omega_+$, $\omega_0$ and $\omega_+$ are the parameters corresponding to infinitesimal
translations, dilatations, and conformal transformations, respectively. The parameter $\lambda$ is called the conformal weight. Further, by expanding the right hand side of (\ref{Phiprime}), we have
\begin{eqnarray}
\delta_{\omega_+} \Phi (t) &=&     \partial_t \Phi(t)  \\
\delta_{\omega_0} \Phi (t) &=&  \left[  t \partial_t  + \lambda  \right] \Phi(t)  \\
\delta_{\omega_-} \Phi (t) &=&   \left[  \ t^2\partial_t \ +2\lambda t \right] \Phi(t) ,
\end{eqnarray}  
where we immediately recognize the generators (\ref{standard-rep}).

The time-derivative of a primary field is not necessarily a primary field. Indeed,
\begin{eqnarray}
\delta_{\omega_+} (\partial_t \Phi) (t) &=&   \left[   \partial_t \ \right] (\partial_t \Phi)(t)  \\
\delta_{\omega_0} (\partial_t \Phi) (t) &=&  \left[   t \partial_t \ + (\lambda +1) \ \right] (\partial_t \Phi)(t) \label{otherrep} \\
\delta_{\omega_-} (\partial_t \Phi) (t) &=&   \left[  t^2\partial_t \ +2 (\lambda +1)t \ + 2 \lambda\  \partial_t ^{-1} \right]  (\partial_t \Phi)(t) .
\end{eqnarray}  

This example motivates us to consider a different representation of $\mathfrak{sl}(2)$, namely
\begin{eqnarray}
 \tilde{E}^+ &=&   -i\partial_t  \nonumber \\
  \tilde{E}^0&=&   i( t \partial_t  + \lambda)  \label{sl2-int-rep}  \\
 \tilde{E}^- &=&   -i( t^2 \ \partial_t  +2 \lambda  t  + \rho  \partial_t ^{-1})  \nonumber \,,
\end{eqnarray}  
where $\rho$ is an additional numerical parameter.
We assume that the functions on that the operators $\tilde E$ act vanish sufficiently fast as
$t\to -\infty$ and define the inverse derivative as an integral
\begin{equation}\label{intrep}
 \partial_t ^{-1} \ f(t) := \int_{-\infty}^{t} \ f(\tau) \ d\tau .
\end{equation} 

To give a name, we shall call the fields transforming under the non-local representation (\ref{sl2-int-rep}) the \emph{hyper-primary} fields, or \emph{hypermary}, for short.

A peculiar feature of the hypermary fields is that they cannot be distinguished from the primary one just looking at the action of $\omega_+$ and $\omega_0$ transformations. However, as we shall see below, the correlation functions differ significantly.

For the hypermary representations, the Casimir operator reads
\begin{equation}
J^2=C(\lambda,\rho)\equiv \lambda (\lambda-1)-\rho\,.\label{Clr}
\end{equation}

To continue our discussion, we shall need the Riemann-Liouville fractional integrals and fractional derivatives \cite{Podnubly1999,Miller-Ross}. The fractional integral of order $u$ reads
\begin{equation}
I^u_x f(x):=\frac 1{\Gamma(u)}\int^x_a f(t)(x-t)^{u -1} dt \,,\label{fint}
\end{equation}
while the fractional derivative of order $s$ is defined as
\begin{equation}
D^{s}_x f(x)=\frac {d^{\left\lceil s \right\rceil}}{dx^{\left\lceil s \right\rceil}}
I^{{\left\lceil s \right\rceil}-s}_xf(x)\,.\label{fdir}
\end{equation}
To ensure consistency with the definition (\ref{intrep}) of the inverse derivative we have to take
$a=-\infty$. 

Of course, the definitions (\ref{fint}) and (\ref{fdir}) make sense only with certain restrictions on the function $f(t)$ and on the parameters $u$ and $s$. We shall not discuss these restrictions here. Exact conditions will be formulated in Sec.\ \ref{sec:3point} where fractional integrals and derivatives will be used to compute the 3-point functions. Also, without any proof or even precise conditions we assume that the following (natural) commutation relations hold\footnote{Obviously, these relation require that $\Phi(t)$ vanishes sufficiently fast at $t\to -\infty$ so that all integral exist and the lower limit of integration does not contribute to the right hand sides of the relations;}
\begin{eqnarray*}
\left[D_t^s, t \right]  \Phi(t) &=& s D_t^{s-1}  \Phi(t),\\
\left[D_t^s, \partial_t \right] \Phi(t) &=& 0,\\
\left[D_t^s, t \partial_t \right] \Phi(t) &=& s D_t^{s}  \Phi(t), \\
\left[D_t^s, t^2 \partial_t \right]  \Phi(t)&=& 2s t D_t^s  \Phi(t)+ s(s-1)D_t^{s-1} \Phi(t)
\end{eqnarray*}
With the help of these relations one can easily show that if $\Phi$ is a hypermary field with the weights $(\lambda, \rho)$ then
\begin{eqnarray*}
\delta_{\omega_+} D_t^s \Phi (t) &=&   \left[ \partial_t \right]  D_t^s \Phi (t)  \\
\delta_{\omega_0} D_t^s \Phi (t) &=&  \left[ t \partial_t + (\lambda +s) \right]  D_t^s \Phi (t) \\
\delta_{\omega_-} D_t^s \Phi (t) &=&   \left[  t^2\partial_t +2 (\lambda +s)t  + (2 s \lambda+s (s-1)+\rho)  \partial_t ^{-1} \right]  D_t^s \Phi (t) ,
\end{eqnarray*}  
i.e., $D^s_t \Phi$ is hypermary with the weights $(\lambda+s, \rho + 2s\lambda + s(s-1))$. The Casimir (\ref{Clr}) is equal for $\Phi$ and $D^s_t \Phi$.

\section{Correlation functions}
%\subsection{Conformal invariance}
The correlation functions of dAFF model have an $\mathfrak{sl}(2)$ invariant form \cite{deAlfaro:1976vlx}, exactly as required by the AdS$_2$/CFT$_1$ correspondence. As was argued in  
\cite{Chamon:2011xk}, this fact is a result of interplay of many non-trivial effects. This may look puzzling since, in particular, the dAFF model has no states that are invariant under all $\mathfrak{sl}(2)$ transformations. However, the desired invariance of correlation functions may be achieved by a suitable modification of the operator-state correspondence rules \cite{Chamon:2011xk}. Since the construction of \cite{Chamon:2011xk} exploits mostly the $SL(2)$ group structure and bears just implicit reference to a particular conformal quantum mechanics, we assume that invariant two- and three-point correlation functions 
\begin{eqnarray}
&&G_2(t_1,t_2)=\langle \phi_{\lambda_1,\rho_1}(t_1)\phi_{\lambda_2,\rho_2}(t_2)\rangle,\nonumber\\
&&G_3(t,t_1,t_2)=\langle \phi_{\lambda_1,\rho_1}(t_1)\phi_{\lambda,\rho}(t)\phi_{\lambda_2,\rho_2}(t_2)\rangle
\end{eqnarray}
may be constructed in some way from the operators $\phi_{\lambda,\rho}$ and a suitable averaging $\langle\dots\rangle$ and study the restrictions imposed on their structure by the invariance under all $\mathfrak{sl}(2)$ generators. For primary fields, such an assumption is usually taken for granted as a part of the definition of CFT$_1$, see e.g. the recent paper \cite{Giombi:2017cqn}. Since fractional integrals/derivatives map primaries to hypermaries, out assumption is satisfied if these integrals/derivatives are well defined. This, in turn, means some restrictions of the allowed hypermary weights, see below.

\subsection{Two-point functions}\label{sec:2point}
For a two-point correlation function $G_2(t_1,t_2)$ the invariance conditions with respect to the $\mathfrak{sl}(2)$ transformations corresponding to the weights $(\lambda_1,\rho_1)$ and $(\lambda_2,\rho_2)$ read
\begin{eqnarray}
(\partial_{t_1}+\partial_{t_2})\ G_2(t_1,t_2) &=& 0,\\
(t_1 \partial_{t_1}+t_2 \partial_{t_2}+ \lambda_1+\lambda_2)\ G_2(t_1,t_2) &=& 0,\\
(t_1^2 \partial_{t_1}+t_2^2 \partial_{t_2}+ 2\lambda_1 t_1+2 \lambda_2 t_2+ \rho_1 \ \partial_{t_1}^{-1}  + \rho_2 \ \partial_{t_2}^{-1})\ G_2(t_1,t_2) &=& 0.
\end{eqnarray}  
The first of the equations above tells us that $G_2(t_1,t_2)$ is a function of the difference $t_1-t_2$, while the second equation fixes this function to
\begin{equation}
G_2(t_1,t_2)\propto \frac{1}{(t_1-t_2)^{\lambda_1+\lambda_2}}\,.\label{G2}
\end{equation} 
The last invariance condition yields the following relation between the weights
\begin{equation}
\lambda_1-\lambda_2 - \frac{\rho_1-\rho_2}{\lambda_1+\lambda_2-1}=0\,.\label{cond3two}
\end{equation}
To be able to apply the inverse derivatives, we also need the condition $\lambda_1+\lambda_2>1$. 

The condition (\ref{cond3two}) can be rewritten as
\begin{equation}
C(\lambda_1,\rho_1)=C(\lambda_2,\rho_2) \,.\label{cond4two}
\end{equation}
We see that in contrast to the case of primary fields, the two-point correlation functions may be non-zero even for different conformal weights.

Two important remarks are in order.
\begin{enumerate}
\item The operators $\partial_{t_1}^{-1}$ and $\partial_{t_2}^{-1}$ if understood as integrals over the real axis are not well defined on the function (\ref{G2}) due to divergences either at infinity or at $t_1=t_2$. To exclude divergences at the infinity, we restrict the weights by the condition
\begin{equation}
\lambda_1+\lambda_2>1\,.\label{l1l2}
\end{equation}
To avoid the singularity at $t_1=t_2$, we shift it to an arbitrary small value up or down in the complex plane and put the branch cut (if needed) so that it does not intersect the real axis. This procedure corresponds to defining the phases in the correlation function which are ambigous in the formula (\ref{G2}).
\item The same expression (\ref{G2}) could have been obtained by taking fractional derivatives or fractional integrals of the standard correlation function with $\rho_1=\rho_2=0$. All comments regarding the singularities and brunch cuts apply also here.
\end{enumerate}

\subsection{Three-point functions}\label{sec:3point}
Here we restrict our attention to the correlation functions of two primary fields with the weights
$(\lambda_1,0)$ and $(\lambda_2,0)$ and one hypermary field with the weight $(\lambda,\rho)$. The method presented below may be extended to a more general situation at the expense of some technical complications.

The $\mathfrak{sl}(2)$ invariance conditions read
\begin{eqnarray}\label{3-point-1}
(\partial_{t_1}+\partial_{t_2}+\partial_{t})\ G_3(t; t_1,t_2) = 0,\\\label{3-point-2}
(t_1 \partial_{t_1}+t_2 \partial_{t_2}+ t\partial_{t}+ \lambda_1+\lambda_2+\lambda)\ G_3(t;t_1,t_2) = 0,\\  
(t_1^2 \partial_{t_1}+t_2^2 \partial_{t_2}+t^2 \partial_{t}+ 2\lambda_1 t_1+2 \lambda_2 t_2+2\lambda t+\rho \ \partial_{t}^{-1})\ G_3(t; t_1,t_2) = 0. \label{3-point-3}
\end{eqnarray}  
Solving these equations directly is too complicated. We proceed in a different way. First, we prove that certain fractional integral of the usual 3-point function with $\rho =0$ satisfies the invariance conditions. Next, we compute this fractional integral.

Let us assume that there exist numbers $\tilde\lambda$ and $s$ such that
\begin{equation}
\lambda = \tilde{\lambda}+s,\qquad
\rho = 2 \tilde{\lambda} s + s(s-1),\label{tillam} 
\end{equation}
i.e., that the hypermary field with the weights $(\lambda,\rho)$ may be obtained (formally) by acting with the fractional derivative of order $s$ on a primary field with the conformal weight $\tilde\lambda$. For $s$, we have 
\begin{equation}
s=\tfrac 12 \bigl( 2\lambda -1 \pm \sqrt{4C(\lambda,\rho)+1} \bigr)
\end{equation}
so that a real solution exists iff $C(\lambda,\rho)\geq -\tfrac 14$. 

The form of conformally invariant 3-point functions of primary fields  with the weights $\tilde\lambda$, $\lambda_1$ and $\lambda_2$ is very well known. Up to a possible constant factor, it reads
\begin{equation}
\tilde G_3(t;t_1,t_2)=(t-t_1)^\alpha (t_1-t_2)^\gamma (t_2-t)^\beta\,,\label{G3prim}
\end{equation}
where
\begin{eqnarray}
&&\alpha =\lambda_2-\lambda_1-\tilde\lambda \,,\nonumber\\
&&\beta = \lambda_1-\lambda_2-\tilde\lambda \,,\label{abgam}\\
&&\gamma=\tilde\lambda -\lambda_1-\lambda_2 \,.\nonumber
\end{eqnarray}

Let us define the function
\begin{equation}
G_3^{v}(t,t_1,t_2)=\int_{-\infty}^{t} d\tau \ f(\tau, t,t_1,t_2)\,,\label{G3v}
\end{equation}
where
\begin{equation}
f(\tau, t,t_1,t_2)=(\tau-t_1)^{\alpha}(t_1-t_2)^{\gamma}(\tau-t_2)^{\beta} (t-\tau)^{v-1}\,.\label{ftau}
\end{equation}
Up to an inessential constant factor this just a fractional integral of (\ref{G3prim}),
\begin{equation}
G_3^{v}(t,t_1,t_2)=\Gamma(v)(-1)^\beta\, I^v_t\tilde G_3(t;t_1,t_2) \,.\label{G3G3}
\end{equation}
To remove the singularities at $\tau=t_1$ and $\tau=t_2$ we give $t_1$ and $t_2$ small imaginary parts and place the brunch cuts in such a way that they do not intersect the real axis. The singularities at the end points of integration $\tau=-\infty$ and $\tau=t$ cannot be removed in this way. We need to impose the following restriction on the parameter to ensure the
convergence of (\ref{G3v})
\begin{equation}
\alpha+\beta+v+1<0,\quad  v>0\,.\label{conv}
\end{equation}
The first of the conditions above is stronger that it is strictly necessary for the convergence of (\ref{G3v}). It also guarantees the existence of $\partial_t^{-1}G_3^v$. At the end of the calculations, $t_1$ and $t_2$ will be returned to the real axis.

Basing on the arguments presented in Sec.\ \ref{sec:rep}, we expect that $G_3^v$ solves the $\mathfrak{sl}(2)$ invariance conditions (\ref{3-point-1}) - (\ref{3-point-3}). A careful analysis is performed in Appendix \ref{sec:appA}. By using the identities derived there we can make the following conclusions.
Eq.\ (\ref{1G3v}) yields the first of the invariance conditions, Eq.\ (\ref{3-point-1}). The second condition, Eq.\ (\ref{3-point-2}), follows from (\ref{2G3v}), (\ref{tillam}), (\ref{abgam}) with the identification
\begin{equation}
v=-s\,.\label{vms}
\end{equation}
The condition (\ref{3-point-3}) follows from (\ref{1401}) and (\ref{1400}). 

We conclude that the $\mathfrak{sl}(2)$ invariance conditions (\ref{3-point-1}) - (\ref{3-point-3}) are indeed satisfied. Next, we have to compute the function $G_3^v(t,t_1,t_2)$. The easiest way to perform this computation is to do the integral when $t$ is smaller than $t_1$ and $t_2$ and then continue the result to the whole range of coordinates. Without any loss of generality we may assume that $t_1<t_2$. The computation of $G_3^v$ for $t<t_1<t_2$ is done in the Appendix \ref{sec:comp}. The result reads:
\begin{eqnarray}
&&G_3^v(t,t_1,t_2)=(-1)^{\alpha+\beta+1}  \frac{\Gamma(1-\alpha-\beta-v)\Gamma(v)}{(\alpha+\beta+v)\Gamma(-\alpha-\beta)} (t_2-t_1)^{\gamma} (t_2-t)^{\alpha+\beta+v} \nonumber \\
&& \ \quad \ \qquad \times \, 
{}_2 F_1\left( -\alpha , -\alpha-\beta-v; -\alpha-\beta;\frac{t_2-t_1}{t_2-t}\right)\nonumber\\
&&=(-1)^{\alpha+\beta+1}  \frac{\Gamma(1-\alpha-\beta-v)\Gamma(v)}{(\alpha+\beta+v)\Gamma(-\alpha-\beta)} (t_2-t_1)^{\gamma} (t_2-t)^{\beta}(t_1-t)^{\alpha+v} \nonumber \\
&& \ \quad \ \qquad \times \, 
{}_2 F_1\left( -\beta , v; -\alpha-\beta;\frac{t_2-t_1}{t_2-t}\right)\,. \label{G3vfin}
\end{eqnarray}
In the last equation we used the identity ${}_2 F_1(a,b;c;z)=(1-z)^{c-a-b}\, {}_2 F_1(c-a,c-b;c;z)$ to rewrite the result in a more symmetric and suggestive way.

The power series (\ref{hyp}) define the hypergeometric function ${}_2 F_1(a,b;c;z)$ for $|z|<1$. This function can be analytically continued to $|z|\geq 1$ along any path which does not contain $z=1$. We take this line to be the real axis for $t$. The values $z=1$ is avoided since we gave complex values to $t_1$ and $t_2$. When the regularizations is removed, i.e., when both $t_1$ and $t_2$ return to the real axis, the correlation function receives a singularity, which is the branching point of the hypergeometric function that we have already mentioned above.

We have computed the 3-point correlation function of two primary fields and one hypermary field. However, some restrictions (\ref{conv}) on the range of parameters were necessary. In particular, the order $v$ of fractional integral was (quite naturally) positive. The function (\ref{G3vfin}) can now be differentiated arbitrary number of times, thus passing from fractional integrals to fractional derivatives, see (\ref{fdir}), and shifting the parameter $s$ in (\ref{tillam}) to positive values. However, the initial conformal weights $\lambda_1$, $\lambda_2$ and $\tilde\lambda$ are restricted by the inequalities (\ref{conv}).

\section{Conclusions}\label{sec:con}
In this note, we suggested a method and computed the 2- and 3-point correlation functions for a conformal quantum mechanics for the fields that transform under a non-local representation of $\mathfrak{sl}(2)$. Our method is based on the invariance argument and used the Riemann-Liouville fractional integrals and fractional derivatives. Interestingly, just one of the generators  is modified by the presence of a non-local operator, so that our "hypermary" field may be easily mistaken for primaries. We expect our results to be useful in various models of the AdS$_2$/CFT$_1$ correspondence. 

Our results may in principal be extended to other minimal realizations of $\mathfrak{sl}(2)$ discussed in \cite{Joseph:1974hr} (that means in this case the realizations of generators of $\mathfrak{sl}(2)$ as rational functions of the canonical pair $t$ and $\partial_t$).

\subsection*{Acknowledgments}
The work of SK was supported by the grant of CNPq 152071/2016-4. The work of DV was supported in part by the grants of CNPq 401180/2014-0 and 303807/2016-4 and by the grant 2016/03319-6 of the S\~ao Paulo Research Foundation (FAPESP). 

\appendix

\section{Identities involving $G_3^v$}\label{sec:appA}
In this Appendix we derive some identities involving the function $G_3^v$ defined by Eq.\ (\ref{G3v}). For the derivatives of $G_3^v$, we have
\begin{eqnarray}
&&\partial_t G_3^{v} 
=\int_{-\infty}^{t} d\tau \ \partial_t f(\tau, t,t_1,t_2) \ + f(\tau, t,t_1,t_2)|_{\tau=t}\nonumber\\
&&\partial_{t_1} G_3^{v} 
=\int_{-\infty}^{t} d\tau \partial_ {t_1} f(\tau, t,t_1,t_2) \nonumber
\end{eqnarray}
and similarly for $\partial_{t_2}G_3^v$. The derivatives of $f$ read
\begin{eqnarray}
&&\partial_t f(\tau, t,t_1,t_2) = \frac{v-1}{t-\tau} f(\tau, t,t_1,t_2),\nonumber\\
&&\partial_{t_1} f(\tau, t,t_1,t_2) = \left(\frac{\gamma}{t_1-t_2}-\frac{\alpha}{\tau-t_1}\right)  f(\tau, t,t_1,t_2),\nonumber\\
&&\partial_{t_2} f(\tau, t,t_1,t_2) = \left(-\frac{\gamma}{t_1-t_2}-\frac{\beta}{\tau-t_2}\right) f(\tau, t,t_1,t_2).\nonumber
\end{eqnarray} 
Therefore,
\begin{eqnarray}
&&\left( \partial_t+\partial_{t_1}+\partial_{t_2}\right) f(\tau, t,t_1,t_2)=  -\partial_\tau f(\tau, t,t_1,t_2)\nonumber\\
&&\left(t \partial_t+ t_1\partial_{t_1}+t_2\partial_{t_2}\right) f(\tau, t,t_1,t_2)
\nonumber\\ && \qquad\qquad
= \bigl(-\tau \partial_\tau+(\alpha+\beta+\gamma +v-1)\bigr) f(\tau, t,t_1,t_2),\nonumber\\
&&\left(t^2 \partial_t+ t_1^2\partial_{t_1}+t_2^2 \partial_{t_2}\right) f(\tau, t,t_1,t_2) \nonumber\\
 &&\qquad\qquad= \bigl( (v-1) t + (\alpha +\gamma ) t_1 + (\beta +\gamma ) t_2\nonumber\\
&&\qquad\qquad\   +(\alpha +\beta +v-1) \tau -\tau^2 \partial_\tau \bigr) f(\tau, t,t_1,t_2)\nonumber
\end{eqnarray}
Consequently,
\begin{eqnarray}\nonumber
&&(\partial_{t}+\partial_{t_1}+\partial_{t_2})\ G_3^v \\&&\qquad = \int_{-\infty}^{t} d\tau \bigl( \partial_t+\partial_{t_1}+\partial_{t_2}\} f(\tau, t,t_1,t_2) 
+ f(\tau, t,t_1,t_2)|_{\tau=t}\nonumber \\
&&\qquad = - \int_{-\infty}^{t} d\tau \ \partial_{\tau} f(\tau, t,t_1,t_2)
+ f(\tau, t,t_1,t_2)|_{\tau=t} \nonumber\\ 
&&\qquad = f(\tau, t,t_1,t_2)|_{\tau=-\infty}
= 0. \label{1G3v}
\end{eqnarray}
One the last line we used the condition (\ref{conv}). Further we have
\begin{eqnarray}\nonumber
&&(t \ \partial_{t}+ t_1 \ \partial_{t_1}+t_2 \ \partial_{t_2})\ G_3^v  \\  \nonumber
&&\qquad = \int_{-\infty}^{t} d\tau \ \bigl(-\tau \partial_\tau+(\alpha+\beta+\gamma +v-1)\bigr)  f(\tau, t,t_1,t_2) \\ \nonumber &&\qquad\ + t \ f(\tau, t,t_1,t_2)|_{\tau=t} \\ \nonumber
&&\qquad =  -\tau \  f(\tau, t,t_1,t_2)|_{\tau= -\infty}^t+(\alpha+\beta+\gamma +v) \int_{-\infty}^{t} d\tau \  f(\tau, t,t_1,t_2) \\ \nonumber
&&\qquad\  +\ t \ f(\tau, t,t_1,t_2)|_{\tau=t} \,.
\end{eqnarray}
This equation yields
\begin{eqnarray}\nonumber
&&(t \partial_{t}+ t_1 \partial_{t_1}+t_2 \partial_{t_2}-(\alpha+\beta+\gamma +v)) G_3^v\\ &&\qquad = -\tau  f(\tau, t,t_1,t_2)|_{\tau= -\infty}=0 \,.\label{2G3v}
\end{eqnarray}
Here we used the conditions (\ref{conv}). Next,
\begin{eqnarray}\nonumber
&&(t^2  \partial_{t}+ t_1^2  \partial_{t_1}+t_2^2  \partial_{t_2}) G_3^v  \\ \nonumber
&&\qquad=\int_{-\infty}^{t} d\tau \  \bigl( (v-1) t+ (\alpha +\gamma ) t_1 + (\beta +\gamma ) t_2\\ \nonumber
&&\qquad\ +(\alpha +\beta +v-1) \tau -\tau^2\partial_\tau \bigr) f(\tau, t,t_1,t_2) 
+ t^2 f(\tau, t,t_1,t_2)|_{\tau=t} \\ \nonumber
&&\qquad=\bigl( (\alpha+\beta+2 v)t+(\alpha+\gamma)t_1+ (\beta+\gamma) t_2 \bigr) \int_{-\infty}^{t} d\tau   f(\tau, t,t_1,t_2)\\ \nonumber
&&\qquad\ - (\alpha+\beta+v+1)  \int_{-\infty}^{t} d\tau \ (t-\tau)   f(\tau, t,t_1,t_2),.\label{1401}
\end{eqnarray}
The last integral is immediately recognized with the help of (\ref{G3G3}) as
\begin{eqnarray}
&&\int_{-\infty}^{t} d\tau \ (t-\tau)   f(\tau, t,t_1,t_2) = G_3^{v+1}=(-1)^\beta\Gamma(v+1)I^{v+1}_t \tilde G_3\nonumber\\
&&\qquad\qquad = (-1)^\beta v\Gamma(v)\partial_t^{-1}I_t^v \tilde G_3
=v\partial^{-1}_t G_3^v. \label{1400}
\end{eqnarray}
Here we used the semigroup property of Riemann-Liouville fractional integrals.
Again, the conditions (\ref{conv}) are sufficient to guarantee the convergence. 

\section{Computation of $G_3^v$}\label{sec:comp}
Let us compute $G_3^v$ in the region $t< t_1 < t_2$. We have
\begin{equation*}
G_3^{v}=(-1)^{\alpha+\beta} (t_2-t_1)^{\gamma} \int_{-\infty}^{t}  (t_1-\tau)^{\alpha} (t_2-\tau)^{\beta} (t-\tau)^{v-1} d\tau\,.
\end{equation*}
In this region, we may use the expansions
\begin{eqnarray*}
(t_1-\tau)^\alpha &=& (t_2-\tau)^\alpha \left(1-\frac{t_2-t_1}{t_2-\tau}\right)^\alpha\\
&=& (t_2-\tau)^\alpha \sum_{n=0}^{\infty} (-1)^n C_{\alpha}^{n} \left (\frac{t_2-t_1}{t_2-\tau}\right)^n\\
&=& (t_2-\tau)^\alpha \sum_{n=0}^{\infty} \frac{(-\alpha)_n}{n!} \left (\frac{t_2-t_1}{t_2-\tau}\right)^n
\end{eqnarray*}
and
\begin{eqnarray*}
(t-\tau)^{v-1} &=& (t_2-\tau)^{v-1} \left(1-\frac{t_2-t}{t_2-\tau}\right)^{v-1}\\
&=& (t_2-\tau)^{v-1} \sum_{m=0}^{\infty} (-1)^m C_{v-1}^{m} \left (\frac{t_2-t}{t_2-\tau}\right)^m\\
&=& (t_2-\tau)^{v-1} \sum_{m=0}^{\infty} \frac{(1-v)_m}{m!} \left (\frac{t_2-t}{t_2-\tau}\right)^m
\end{eqnarray*}
Here
\begin{equation*}
C_{p}^{q} = \frac{\Gamma(p+1)}{\Gamma(q+1)\Gamma(p-q+1)}
\end{equation*}
are binomial coefficients. In particular, for $n\in \mathbb{N}$,
\begin{equation*}
C_{\alpha}^{n} = (-1)^n \frac{(-\alpha)_n}{n!}=\frac{\alpha(\alpha-1)\cdots (\alpha-n+1)}{n!}\,,
\end{equation*}
where $(p)_n= p(p+1)\cdots(p+n-1)=\frac{\Gamma(p+n)}{\Gamma(p)}$ is the increasing Pochhammer symbol.
Then,
\begin{eqnarray*}
&&G_3^{v} 
= (-1)^{\alpha+\beta} (t_2-t_1)^{\gamma} \sum_{m,n}  \frac{(-\alpha)_n}{n!}  \frac{(1-v)_m}{m!} (t_2-t_1)^n(t_2-t)^m  \\
&&\qquad \times\int_{-\infty}^{t} (t_2-\tau)^{\alpha+\beta+v-1-m-n} d\tau\\
&&\quad =    (-1)^{\alpha+\beta+1} (t_2-t_1)^{\gamma} \sum_{m,n}  \frac{(-\alpha)_n}{n!}  \frac{(1-v)_m}{m!}  
 \frac{(t_2-t_1)^n(t_2-t)^{\alpha+\beta+v-n}}{\alpha+\beta+v-m-n}
\\
&&\quad=(-1)^{\alpha+\beta+1} (t_2-t_1)^{\gamma} (t_2-t)^{\alpha+\beta+v} \sum_{n}  \frac{(-\alpha)_n}{n!} \left(\frac{t_2-t_1}{t_2-t}\right)^n \\
&&\qquad \times \sum_{m} \frac{(1-v)_m}{m!}
 \frac{1}{\alpha+\beta+v-m-n}\,.
\end{eqnarray*}
The sum on the last line equals to
\begin{equation*}
\sum_{m} \frac{(1-v)_m}{m!(\alpha+\beta+v-m-n)} =\frac{\Gamma(1-\alpha-\beta-v+n)\Gamma(v)}{(\alpha+\beta+v-n)\Gamma(n-\alpha-\beta)}
\end{equation*}
The sum over $n$ is performed by using the hypergeometric series,
\begin{equation}
\sum_{n=0}^\infty \frac{(a)_n(b)_n}{(c)_n\, n!} \, z^n={}_2 F_1(a,b;c;z)\,,\label{hyp}
\end{equation}
where ${}_2 F_1$ is the hypergeometric function, $|z|<1$. By collecting everything together we arrive at the expression (\ref{G3vfin}) in the main text of this article.


\begin{thebibliography}{99}
\bibitem{deAlfaro:1976vlx} 
  V.~de Alfaro, S.~Fubini and G.~Furlan,
  %``Conformal Invariance in Quantum Mechanics,''
  Nuovo Cim.\ A {\bf 34}, 569 (1976).
  doi:10.1007/BF02785666
  %%CITATION = doi:10.1007/BF02785666;%%
	
\bibitem{Jackiw:1972cb} 
  R.~Jackiw,
  %``Introducing scale symmetry,''
  Phys.\ Today {\bf 25N1}, 23 (1972)
  %[Phys.\ Today {\bf 25}, 23 (1972)].
  doi:10.1063/1.3070673
  %%CITATION = doi:10.1063/1.3070673;%%

\bibitem{Chamon:2011xk} 
  C.~Chamon, R.~Jackiw, S.~Y.~Pi and L.~Santos,
  %``Conformal quantum mechanics as the CFT$_1$ dual to AdS$_2$,''
  Phys.\ Lett.\ B {\bf 701}, 503 (2011)
  doi:10.1016/j.physletb.2011.06.023
  [arXiv:1106.0726 [hep-th]].
  %%CITATION = doi:10.1016/j.physletb.2011.06.023;%%

\bibitem{Jackiw:2012ur} 
  R.~Jackiw and S.-Y.~Pi,
  %``Conformal Blocks for the 4-Point Function in Conformal Quantum Mechanics,''
  Phys.\ Rev.\ D {\bf 86}, 045017 (2012)
  Erratum: [Phys.\ Rev.\ D {\bf 86}, 089905 (2012)]
  doi:10.1103/PhysRevD.86.045017, 10.1103/PhysRevD.86.089905
  [arXiv:1205.0443 [hep-th]].
	
\bibitem{Fedoruk:2011aa}
  S.~Fedoruk, E.~Ivanov and O.~Lechtenfeld,
  %``Superconformal Mechanics,''
  J.\ Phys.\ A {\bf 45} (2012) 173001
  doi:10.1088/1751-8113/45/17/173001
  [arXiv:1112.1947 [hep-th]].
  %%CITATION = doi:10.1088/1751-8113/45/17/173001;%%
	
\bibitem{Khodaee:2012bg}
  S.~Khodaee and F.~Toppan,
  %``Critical scaling dimension of D-module representations of N=4,7,8 Superconformal Algebras and constraints on Superconformal Mechanics,''
  J.\ Math.\ Phys.\  {\bf 53} (2012) 103518
  doi:10.1063/1.4758923
  [arXiv:1208.3612 [hep-th]].
  %%CITATION = doi:10.1063/1.4758923;%%

\bibitem{Vassilevich:2013ai} 
  D.~V.~Vassilevich,
  %``Holographic duals to poisson sigma models and noncommutative quantum mechanics,''
  Phys.\ Rev.\ D {\bf 87}, no. 10, 104011 (2013)
  doi:10.1103/PhysRevD.87.104011
  [arXiv:1301.7029 [hep-th]].
  %%CITATION = doi:10.1103/PhysRevD.87.104011;%%

\bibitem{Maldacena:2016upp} 
  J.~Maldacena, D.~Stanford and Z.~Yang,
  %``Conformal symmetry and its breaking in two dimensional Nearly Anti-de-Sitter space,''
  PTEP {\bf 2016}, no. 12, 12C104 (2016)
  doi:10.1093/ptep/ptw124
  [arXiv:1606.01857 [hep-th]].
  %%CITATION = doi:10.1093/ptep/ptw124;%%

\bibitem{Podnubly1999} 
  I. Podlubny,
  %``Fractional Differential Equations''
 Fractional Differential Equations, Academic Press, San Diego, (1999)
ISBN: 978-0471588849
    
  %\cite{Miller-Ross}
\bibitem{Miller-Ross} 
  K. S. Miller and B. Ross,
  %``An Introduction to the Fractional Calculus and Fractional Differential Equations''
 An Introduction to the Fractional Calculus and Fractional Differential Equations, Wiley-Interscience, NY (1993)
 ISBN: 0-12 558840-2
	
\bibitem{Giombi:2017cqn} 
  S.~Giombi, R.~Roiban and A.~A.~Tseytlin,
  ``Half-BPS Wilson loop and AdS$_2$/CFT$_1$,''
  arXiv:1706.00756 [hep-th].
  %%CITATION = ARXIV:1706.00756;%%
	
\bibitem{Joseph:1974hr} 
  A.~Joseph,
  %``Minimal realizations and spectrum generating algebras,''
  Commun.\ Math.\ Phys.\  {\bf 36}, 325 (1974).
  doi:10.1007/BF01646204
  %%CITATION = doi:10.1007/BF01646204;%%
\end{thebibliography}
\end{document}